\documentclass[twocolumn,showpacs,preprintnumbers,amsmath,amssymb,prb]{revtex4-1}
\usepackage{stmaryrd}
\usepackage{mathrsfs}
\usepackage{graphicx}
\usepackage{dcolumn}
\usepackage{bm}
\usepackage{amsmath}
\usepackage{amsfonts}

\begin{document}

\title{Two-dimensional time-reversal-invariant topological superconductivity in a doped quantum spin Hall insulator}
\author{Jing Wang}
\author{Yong Xu}
\author{Shou-Cheng Zhang}
\affiliation{Department of Physics, McCullough Building, Stanford University, Stanford, California 94305-4045, USA}

\begin{abstract}
Time-reversal-invariant topological superconductors have a full paring gap in the bulk and gapless Majorana states at
the edge or on the surface. Here, we theoretically propose topological superconductivity in a doped quantum spin
Hall insulator. We study the pairing symmetry of a doped two-dimensional tin film within a two-orbital model, and
find that a novel spin-triplet pairing is favored when the interorbital attractive interaction is stronger
than the intraorbital interaction. We propose that a doped tin film is a good candidate for
a $2d$ topological superconductor. Edge channels are studied in a tight-binding model numerically. Finally, we discuss
the robustness of topological superconductivity in two-dimensional tin films by comparing to $3d$ superconductivity in bulk tin.
\end{abstract}

\date{\today}


\maketitle

\section{Introduction}

The search for topological states of quantum matter has generated intensive interest in condensed matter physics~\cite{qi2010a,hasan2010,qi2011,moore2010}.
Recently, the quantum spin Hall (QSH) state in two dimensions (2D) and topological insulators in three dimensions (3D) have been theoretically predicted and experimentally observed in a number of materials~\cite{bernevig2006c,koenig2007,zhang2009,xia2009}, and both of them are characterized by the $Z_2$ topological indices~\cite{kane2005,fu2007,qi2008}. Soon afterwards, the concept of time-reversal-invariant (TRI) topological superconductors has been proposed~\cite{qi2009,schnyder2008,kitaev2009}. Closely related to QSH state and topological insulators, the $2d$ and $3d$ TRI topological superconducting state has a full pairing gap in the bulk, and gapless Majorana states at the edge and on the surface, respectively, which have potential applications in fault-tolerant topological quantum computation~\cite{nayak2008}. Moreover, an emergent supersymmetry is naturally present in these systems as a consequence of the time-reversal symmetry~\cite{qi2009}. Great efforts have been made to search for topological superconductors, however, finding candidate materials for these new topological phases of matter is still challenging.

A simple and general criterion has been proposed to test for TRI topological superconductor based on the pairing amplitude on the Fermi surface~\cite{qi2010}. A $2d$ TRI superconductor is nontrivial if there are an \emph{odd} number of Fermi surfaces with a \emph{negative} pairing order parameter~\cite{qi2010}. This physical criteria suggests to search for topological superconductors in nonconventional superconducting materials with inversion symmetry breaking~\cite{qi2010} and strong correlation~\cite{scalapino1986}. Recently, superconductivity has been realized in a doped topological insulator Cu$_x$Bi$_2$Se$_3$~\cite{hor2010}. Such material has been proposed to be a $3d$ topological superconductor, where a novel spin-triplet pairing with odd parity is favored by strong spin-orbit coupling (SOC) based on a two-orbital model~\cite{fu2010}. However, the pairing symmetry in doped Bi$_2$Se$_3$ is still under active debates~\cite{sasaki2011,kirzhner2012,bay2012,levy2013}. On the other hand, the $2d$ TRI topological superconductor has not yet been discovered. There are some theoretical discussions on possible TRI topological superconductivity in noncentrosymmetric superconductors with the Rashba spin splitting~\cite{tanaka2009,sato2009a,nakosai2012}. The key point here is for the spin split bands, one is paired into ($p_x+ip_y$) state, the other is paired into ($p_x-ip_y$) state. However, to realize TRI topological superconductor, the spin triplet $p$-wave pairing should be dominant over spin singlet $s$-wave pairing in the two spin split bands~\cite{qi2009,tanaka2009}.

Doped band insulators with strong SOC may be good candidate materials in realizing an exotic pairing~\cite{wan2014}. QSH effect has been realized in heterostructures~\cite{qi2011}, these systems have the advantage of great controllability of structure, doping, symmetry, and SOC. In this paper, we theoretically propose topological superconductivity in a doped QSH insulator. We study the pairing symmetry of the newly predicted QSH insulator in decorated stanene films Sn$X$ ($X=$ -OH, -F, -Cl, -Br, and -I)~\cite{xu2013} within a two-orbital model for its band structure. When the interorbital attractive interaction is stronger than the intraorbital interaction, the resulting state realizes a topological superconductor. We explicitly calculate the Majorana edge spectrum in a tight-binding model. Finally, we discuss the robustness of topological superconductivity in doped Sn$X$ films by comparing to $3d$ superconductivity in bulk $\beta$-tin.

The organization of this paper is as follows. After this
introductory section, Sec. II describes the effective two-orbital model
for the superconductivity in a doped QSH insulator. Section III presents the
results on the pairing symmetry, phase diagram and edge state. Section IV presents discussion
and possible experimental realization of topological superconductivity in tin film.

\section{Model}

\begin{figure*}[t]
\begin{center}
\includegraphics[width=6.9in]{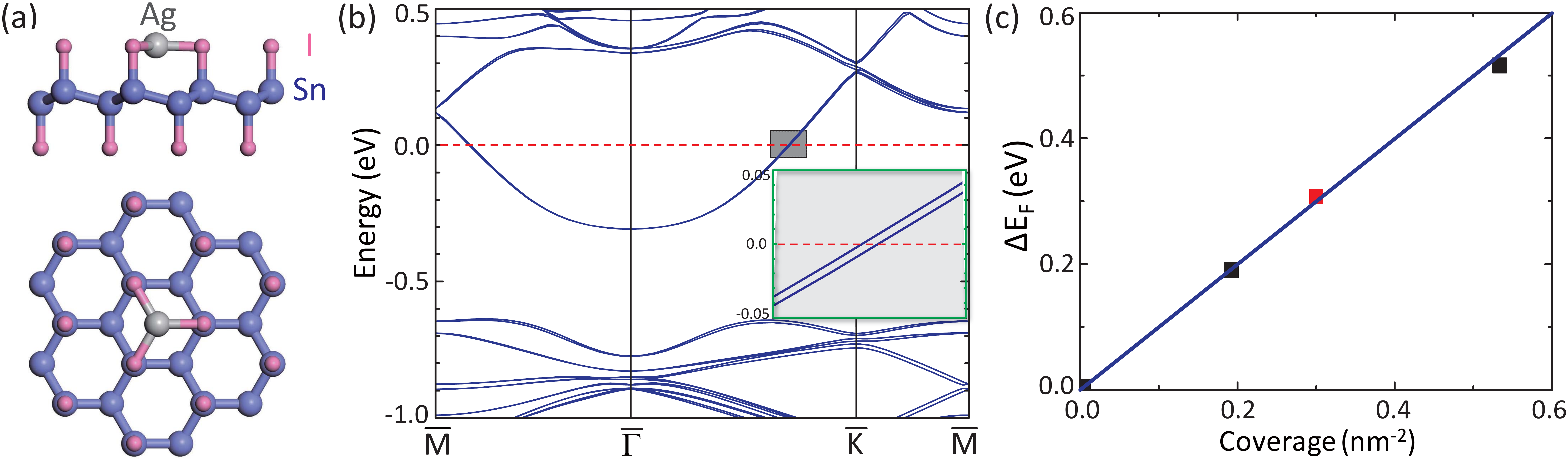}
\end{center}
\caption{(color online) (a) Crystal structure for SnI film with Ag adatom from the side (top) view [upper
(lower)]. Inversion symmetry is broken due to Ag doping. (b) First-principles calculations of band structure for SnI with one Ag adatom on a $4\times4$ surface supercell. The Fermi level is indicated by the dashed line. Inset shows the spin splitting. (c) Fermi energy vs Ag doping concentration.}
\label{fig1}
\end{figure*}

Study of superconductivity in a doped QSH insulator requires the knowledge of its band structure and pairing mechanism. The general results presented in this paper is generic for any doped QSH insulators. Here, we would like to start from a simple model describing the 2D QSH insulator Sn$X$ for concreteness~\cite{xu2013}. In fact, 3D $\beta$-Sn was one of the first superconductors to be studied experimentally, with the critical transition temperature $3.72$~K. Therefore, it is likely that doped Sn$X$ film is also a superconductor. In the following, we assume that 2D Sn$X$ is superconducting, and study under what condition it would also be a $2d$ TRI topological superconductor. We leave the discussion on pairing mechanism to the end.

A low-buckled geometry for 2D SnI is shown in Fig.~\ref{fig1}(a), where they have a stable $sp^3$ configuration analogous to graphane. The lattice symmetry is $D_{3d}$. As shown in Fig.~\ref{fig1}(b) by first-principles calculations, the band structure of Ag$_x$SnI is similar to its parent compound SnI. The low-energy bands of SnI consist of an antibonding state of $s$ orbital and a bonding state of $p_{x,y}$ orbitals, labeled by $|s^-,\uparrow(\downarrow)\rangle$ and $|p_{x,y}^+,\uparrow(\downarrow)\rangle$, respectively, which is similar as for HgTe quantum well~\cite{bernevig2006c}. The effective Hamiltonian describing these four bands near the $\Gamma$ point is given by the model of Bernevig, Hughes and Zhang~\cite{bernevig2006c}:
\begin{equation}\label{model}
\mathcal{H}_0(\mathbf{k}) = \varepsilon(k)+M(k)1\otimes\sigma_3+A(k_xs_3\otimes\sigma_1-k_y1\otimes\sigma_2),
\end{equation}
here, $s_i$ and $\sigma_i$ ($i=1, 2, 3$) are Pauli matrices acting spin and orbital, respectively. To the lowest order in $k$, $M(k)=M_0+M_1(k_x^2+k_y^2)$ and $\varepsilon(k)=D_0+D_1(k_x^2+k_y^2)$ account for the particle-hole asymmetry~\cite{parameters}. $M_0>0$ and $M_1<0$ guarantee that the system is in the inverted regime. The basis of Eq.~(\ref{model}) is $|s^-,\uparrow\rangle$, $|p^+_{x,y},\uparrow\rangle$, $|s^-,\downarrow\rangle$, $|p^+_{x,y},\downarrow\rangle$, and the $\pm$ in the basis stand for the even and odd parity and $\uparrow$, $\downarrow$ represent spin up and down states, respectively.

With the chemical doping by Ag adatom or electrical gating, the lattice symmetry is reduced to $D_3$. Therefore, additional Rashba terms will be added into effective Hamiltonian due to inversion symmetry breaking~\cite{rothe2010}. To the lowest order, the only possible term is
\begin{equation}
\mathcal{H}_{R}(\mathbf{k})= \alpha\left(s_2k_x-s_1k_y\right)\otimes(\sigma_3+1),
\end{equation}
where $\alpha$ determines the strength of spin splitting. The effective model for Ag$_x$SnI is given by $\mathcal{H}=\mathcal{H}_0+\mathcal{H}_R$. The band structure is plotted in Fig.~\ref{fig1}(b), and the bands show small spin splitting which can be tuned to be large by gating. Because of Ag doping, the Fermi energy $\mu$ lies in the conduction band approximately 0.31~eV above the band edge, which leads to a small Fermi surface respecting full rotation symmetry around the $z$ axis. Fig.~\ref{fig1}(c) shows the linear relation between chemical potential and doping concentration, matches well with the parabolic band structure in 2D.

As for the fermion pairing, we consider the following short-range density-density interactions,
\begin{equation}
\mathcal{H}_{\text{int}}(\mathbf{x})=-U\left[n_1^2(\mathbf{x})+n_2^2(\mathbf{x})\right]-2Vn_1(\mathbf{x})n_2(\mathbf{x}),
\end{equation}
where $n_{\sigma=1,2}(\mathbf{x})=\sum_{s=\uparrow,\downarrow}c^{\dag}_{\sigma s}(\mathbf{x})c_{\sigma s}(\mathbf{x})$ is the electron density in orbital $\sigma$. $\sigma=1,2$ represent $s^-$ and $p^+_{x,y}$, respectively. $U$ and $V$ are effective intraorbital and interorbital interaction, respectively. We assume that at least one of them is attractive, say due to phonons as in the case of superconductivity of 3D tin. The two-orbital $U$-$V$ model for 2D Ag$_x$SnI is
\begin{align}\label{uvmodel}
H &= \int d\mathbf{k} c^{\dag}_{\mathbf{k}}\left[\mathcal{H}(\mathbf{k})-\mu\right]c_{\mathbf{k}}+\int d\mathbf{r}\mathcal{H}_{\text{int}}(\mathbf{r}).
\end{align}
In the following, we shall apply the criterion of Ref.~\onlinecite{qi2010} to investigate topological superconductivity in this noncentrosymmetric model.

\section{Results}
\subsection{Pairing symmetry}

To determine the superconducting phase diagram of the $U$-$V$ model, we construct the Bogoliubov-de Gennes (BdG) Hamiltonian with mean-field approximation
\begin{equation}\label{BdG}
H_{\text{BdG}} = \int d\mathbf{k}\Xi^{\dag}_{\mathbf{k}}\left[
\left(\mathcal{H}(\mathbf{k})-\mu\right)\tau_3+\Delta(\mathbf{k})\tau_1\right]\Xi_{\mathbf{k}}.
\end{equation}
Here $\tau_{x,z}$ are Pauli matrices in Nambu space and the basis $\Xi^{\dag}_{\mathbf{k}}\equiv$($c^{\dag}_{1\mathbf{k}\uparrow},c^{\dag}_{2\mathbf{k}\uparrow},c^{\dag}_{1\mathbf{k}\downarrow},
c^{\dag}_{2\mathbf{k}\downarrow},c_{1-\mathbf{k}\downarrow},c_{2-\mathbf{k}\downarrow},-c_{1-\mathbf{k}\uparrow},-c_{2-\mathbf{k}\uparrow}$).
The low energy physics with a small Fermi surface has full rotation symmetry around the $z$ axis $\mathcal{R}_z$ instead of 3-fold rotation symmetry $C_3$ of the point group $D_3$ in the lattice. We classify all possible pairing potential $\Delta(\mathbf{k})$ according to time-reversal symmetry $\mathcal{T}\equiv(is_2\otimes1)K$ with $K$ complex conjugation, and $\mathcal{R}_z=e^{i(\theta/2)\Sigma_z}$ with $\Sigma_z=s_3\otimes(2-\sigma_3)$. In the weak coupling limit with purely short-range interaction, the mean-field pairing potential is $\mathbf{k}$ independent. In Table~\ref{table1}, only 6 forms can have nonzero values among the 16 possible products of $(1,s_1,s_2,s_3)$ and $(1,\sigma_1,\sigma_2,\sigma_3)$. We find three different pairing symmetries with angular momentum $l_z=0, 1, 2$ under $\mathcal{R}_z$. The form of the corresponding pairing order parameter $\Delta_i$, $i=1,2,3$ is shown explicitly
\begin{eqnarray}
\Delta_1:&&\ \ c_{1\uparrow}c_{1\downarrow}+c_{2\uparrow}c_{2\downarrow},c_{1\uparrow}c_{1\downarrow}-c_{2\uparrow}c_{2\downarrow},
\nonumber
\\
\Delta_2:&&\ \ \left(i(c_{1\uparrow}c_{2\downarrow}+c_{1\downarrow}c_{2\uparrow}),c_{1\uparrow}c_{2\downarrow}-c_{1\downarrow}c_{2\uparrow}\right),
\\
\Delta_3:&&\ \ \left(c_{1\uparrow}c_{2\uparrow}+c_{1\downarrow}c_{2\downarrow}, i(c_{1\uparrow}c_{2\uparrow}-c_{1\downarrow}c_{2\downarrow})\right).
\nonumber
\end{eqnarray}
$\Delta_1$ is a spin-singlet, whereas $\Delta_2$ and $\Delta_3$ are interorbital spin-triplets. The symmetry properties of $\Delta_i$ are shown in Table~\ref{table1}.

\begin{table}[t]
\caption{Three possible nonvanishing pairing potentials of the two-orbital $U$-$V$ model, $\Delta_1$, $\Delta_2$, and $\Delta_3$.
Matrix representation are off-diagonal elements of BdG Hamiltonian.}
\begin{center}\label{table1}
\renewcommand{\arraystretch}{1.4}
\begin{tabular*}{3.3in}
{@{\extracolsep{\fill}}cccc}
\hline
\hline
$\Delta$ & Matrix & $\mathcal{R}_z$ & $\mathcal{T}$
\\
\hline
$\Delta_1$ & $1\otimes1$, $1\otimes\sigma_3$ & $0$ & $+$
\\
$\Delta_2$ & $(s_3\otimes\sigma_2,1\otimes\sigma_1)$ & $1$ & $+$
\\
$\Delta_3$ & $(s_1\otimes\sigma_2,s_2\otimes\sigma_2)$ & $2$ & $+$
\\
\hline
\hline
\end{tabular*}
\end{center}
\end{table}

\subsection{Phase diagram}

\begin{figure}[t]
\begin{center}
\includegraphics[width=3.3in]{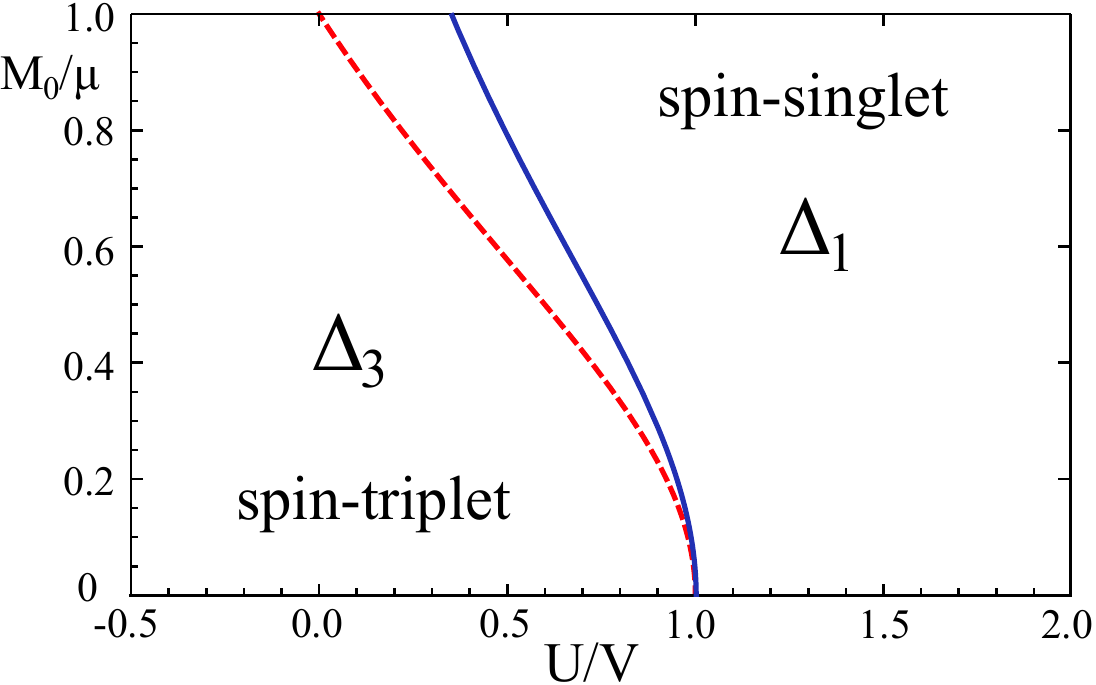}
\end{center}
\caption{(color online) Phase diagram of superconductivity in the two-orbital $U$-$V$ model, showing highest $T_c$ phase as a function of $M_0/\mu$ and $U/V$. $\Delta_2$ cannot be leading instability in this phase diagram, for all the pairing states in $\Delta_{1,3}$ are fully gapped.
Solid (blue) and dashed (red) line are phase boundary for $\alpha\neq0$ and $\alpha=0$, respectively.}
\label{fig2}
\end{figure}

The excitation energy of quasiparticle are obtained by diagonalizing the BdG Hamiltonian Eq.~(\ref{BdG}) with fixing the pairing potential to each $\Delta_i$. We find superconducting gap for $\Delta_2$ has point nodes (in the $k_x$-direction when one choose $s_3\otimes\sigma_2$), and the others have full gap. To obtain the phase diagram, we estimate the superconducting critical temperature $T_c$ by analyzing superconducting susceptibility for each pairing potentials. The standard pairing susceptibility $\chi_0$ is defined as $\chi_0=-T\sum_{\mathbf{k},n}\mathrm{Tr}[\tau_1G(\mathbf{k},i\omega_n)\tau_1G(\mathbf{k},i\omega_n)]$, with $G(\mathbf{k},i\omega_n)=[i\omega_n-(\mathcal{H}(\mathbf{k})-\mu)\tau_3]^{-1}$ the Matsubara Green function. All other susceptibilities $\chi_1$, $\chi_2$, and $\chi_3$ can be obtained by replacing $\tau_1$ with their corresponding pairing potential $\tau_11\otimes\sigma_3$, $\tau_1s_3\otimes\sigma_2$ (or $\tau_11\otimes\sigma_1$), and $\tau_1s_1\otimes\sigma_2$ (or $\tau_1s_2\otimes\sigma_2$) in Table~\ref{table1}. A straightforward calculation shows that they can be expressed by $\chi_0$, which contains the logarithmic divergence at the Fermi surface. The linearized gap equations for $T_c$ in each pairing channel are as
\begin{equation}\label{gapequation}
\begin{aligned}
\Delta_1:&\ \ \det
\left[U
  \begin{pmatrix}
  \chi_0(T_c) & \chi_{01}(T_c)\\
  \chi_{10}(T_c) & \chi_1(T_c)
  \end{pmatrix}
  -1
\right]=0,\\
\Delta_{2,3}:&\ \ V\chi_{2,3}(T_c)=1.
\end{aligned}
\end{equation}
Using the band structure of $\mathcal{H}$, we can calculate the phase diagram numerically. In the limit of $\alpha\rightarrow0$, we obtain the values of $\chi$'s analytically: $\chi_0=\int d\xi D(\xi)\tanh(\xi/2T)/2\xi$, $D(\xi)$ is the density of states. $\chi_{01}=\chi_{10}=(M_0/\mu)\chi_0$, $\chi_1=(M_0/\mu)^2\chi_0$, $\chi_3=2\chi_2=[1-(M_0/\mu)^2]\chi_0$. Because $\chi_2<\chi_3$, we find that $\Delta_2$ always has a lower $T_c$ than $\Delta_3$. From the highest $T_c$, only $\Delta_1$ and $\Delta_3$ appear in the phase diagram. By calculating their $T_c$'s from (\ref{gapequation}), we obtain the phase boundary
\begin{equation}
\frac{U}{V}=\frac{1-(M_0/\mu)^2}{1+(M_0/\mu)^2}.
\end{equation}
Fig.~\ref{fig2} shows the phase diagram as a function $U/V$ and $M_0/\mu$, for positive (attractive) $V$. A significant part of the phase diagram is the $\Delta_3$ phase, especially for the inversion symmetry breaking $\alpha\neq0$.

\subsection{Criterion}

Next we consider the topological nature of the pairing state. The present system belongs to the symmetry class $\mathrm{DIII}$ in 2D~\cite{schnyder2008,kitaev2009}, which is characterized by a $Z_2$ topological invariant in contrast to $Z$ in 3D.
Since the system has $\mathcal{T}$ symmetry without inversion symmetry, the criteria in Ref.~\onlinecite{qi2010} can be applied.
Namely, a TRI superconductor is a topological superconductor if (1) it has a full superconducting gap \emph{and} (2) there are an odd number of Fermi surfaces each of which encloses one TRI point (which satisfy $\Gamma_a=-\Gamma_a$ up to a reciprocal lattice vector) in the Brillouin zone and has negative pairing. The $Z_2$ invariant is
\begin{align}\label{criterion}
\mathcal{N}_{\text{2D}} &= \prod_j\left[\mathrm{sgn}(\delta_j)\right]^{m_j}.
\end{align}
Here $j$ labels the Fermi surface, $m_j$ is the number of the TRI points enclosed by the $j$th Fermi surface, $\mathrm{sgn}(\delta_j)\equiv\mathrm{sgn}[\langle j,\mathbf{k}|\mathcal{T}\Delta^{\dag}|j,\mathbf{k}\rangle]$ denotes the sign of the paring amplitude of the $j$th Fermi surface, $|j,\mathbf{k}\rangle$ is the eigenvectors of $\mathcal{H}(\mathbf{k})$. Here in our system, the $\Delta_3$ pairing has opposite sign on the two Fermi surfaces which gives rise to a topological superconductor phase, while $\Delta_1$ has the same sign as shown in Fig.~\ref{fig3}. If we take the limit $\alpha\rightarrow0$, the two Fermi surfaces become degenerate, still the only odd-parity pairing $\Delta_3$ phase is topological~\cite{fu2010,sato2009b,sato2010}. As shown in Fig.~\ref{fig2}, the topological superconductor phase $\Delta_3$ is more favorable when inversion symmetry is breaking.

\subsection{Edge state}

\begin{figure}[t]
\begin{center}
\includegraphics[width=3.3in]{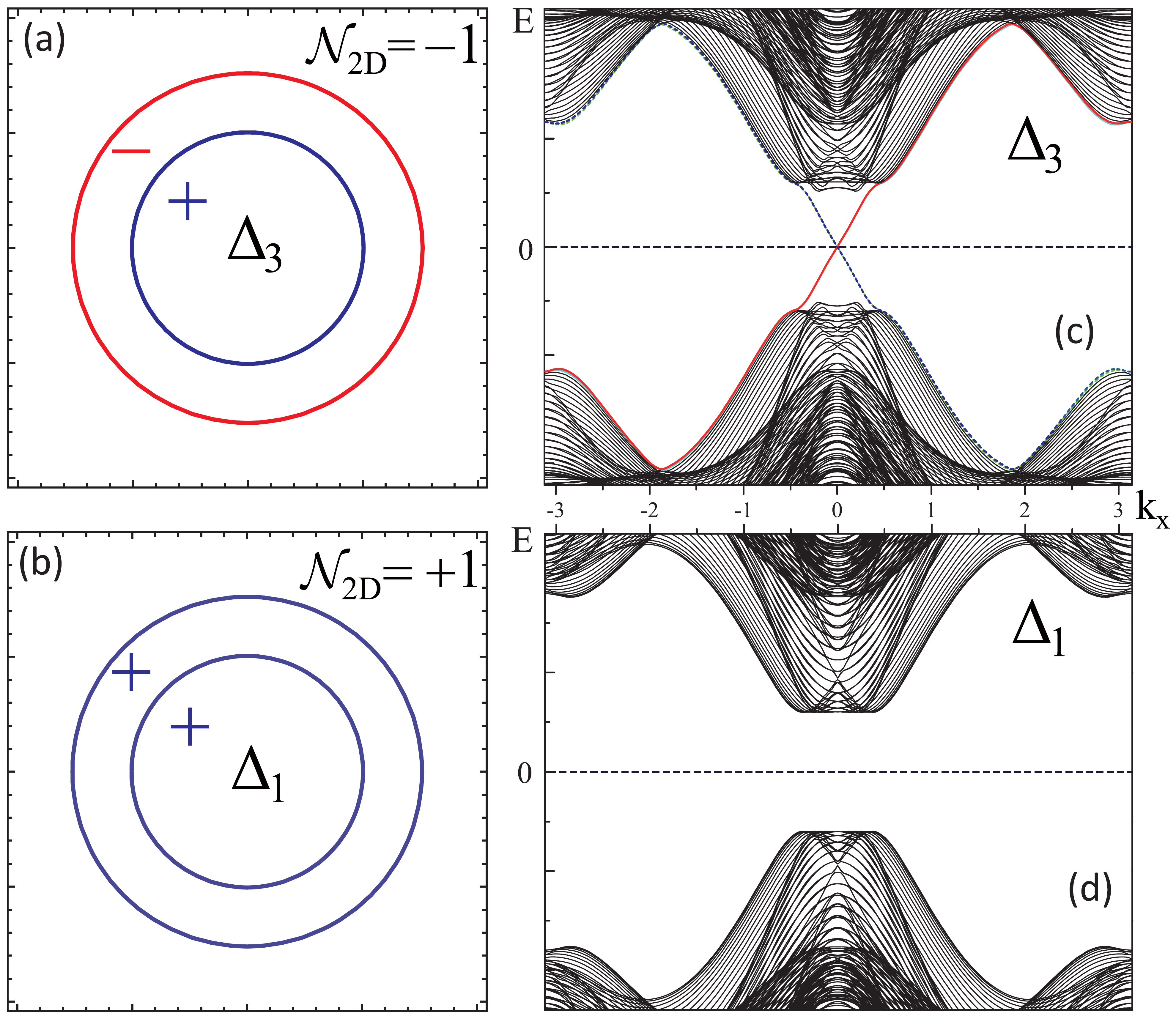}
\end{center}
\caption{(color online) Bulk $Z_2$ topological number, Fermi surface pairing amplitude and edge energy spectrum. (a) $\mathcal{N}_{\mathrm{2D}}=-1$ for $\Delta_3$ pairing, and (c) edge spectrum shows that helical edge modes appears at each edge of the sample in the superconducting gap; while (b) $\mathcal{N}_{\mathrm{2D}}=+1$ for $\Delta_1$ pairing, and (d) no edge states. All parameters are taken from Ref.~\onlinecite{parameters}.}
\label{fig3}
\end{figure}

We confirm the system is exactly in the topological phase under such conditions. To obtain the topological protected gapless edge states, we solve the following tight-binding model describing continuous model Eq.~(\ref{BdG}) in the low energy regime,
\begin{eqnarray}
H &=& \sum\limits_{\langle\mathbf{r}\mathbf{r}'\rangle}c^{\dag}_{\mathbf{r}}t_{\mathbf{rr}'}c_{\mathbf{r}'}-\sum\limits_{\mathbf{r}}\mu'c^{\dag}_{\mathbf{r}}c_{\mathbf{r}}
+\sum\limits_{\mathbf{r}}\left[c^{\dag}_{\mathbf{r}}\Delta c^{\dag}_{\mathbf{r}}-\mathrm{H.c.}\right],
\nonumber
\end{eqnarray}
where $\langle\mathbf{r}\mathbf{r}'\rangle$ denotes the nearest-neighbor site. The hopping parameters $\mu'=\mu-(D_0+4D_1)-(M_0+4M_1)1\otimes\sigma_3$, $t_{\mathbf{r}\mathbf{r}\pm a\hat{\mathbf{x}}}=-(D_1+M_11\otimes\sigma_3)\pm (i/2)[As_3\otimes\sigma_1+\alpha s_2\otimes(\sigma_3+1)]$, and $t_{\mathbf{r}\mathbf{r}\pm a\hat{\mathbf{y}}}=-(D_1+M_11\otimes\sigma_3)\mp (i/2)[A1\otimes\sigma_2+\alpha s_1\otimes(\sigma_3+1)]$. We consider the $\Delta_3$ pairing state in the cylindrical geometry with periodic boundary condition in the $x$ direction and open one in the $y$ direction. The energy spectrum of this model is shown in Fig.~\ref{fig3}(c). One can see that there are helical Majorana states crossing the bulk superconducting gap, where the right-going and left-going states are spin splitting (very small), and localized at opposite edges. Therefore, nontrivial $Z_2$ number in the bulk will lead to helical states at the edge. However, there are no edge states with $\Delta_1$ pairing as in Fig.~\ref{fig3}(d), which are consistent with the previous study on bulk topological invariant.

\section{Discussion}

Finally, we discuss the robustness of topological superconductivity obtained from the two-orbital $U$-$V$ model and the possible pairing mechanism.
As the phase diagram shows, the \emph{interorbital} spin-triplet $\Delta_3$ phase wins as long as the interorbital interaction exceeds over the
intraorbital one ($V>U$). This arises from the specific form of SOC in the band structure, which favors $\Delta_3$ pairing. Also, inversion symmetry breaking and \emph{multi}-orbital system would result in more unconventional pairing. Therefore, doped QSH insulators with strong SOC would offer better way to find topological superconductors. The realistic value of $U$ and $V$ in tin films are difficult to estimate. Nonetheless, the $3d$ superconductivity in bulk $\beta$-Sn is $s$-wave pairing from $s$ orbit via phonon-mediated mechanism. It is likely that 2D Sn$X$ is also a superconductor with a phonon driven pairing mechanism. The bare value of attractive $U_{\text{ph}}$ and $V_{\text{ph}}$ are given by the electron-phonon coupling $\lambda_{\text{2d}}$ as $U_{\text{ph}}, V_{\text{ph}}\propto\lambda^2_{\text{2d}}$. Such electron-phonon coupling strength $\lambda_{\text{2d}}$ in 2D can be modulated and could even be larger than that in 3D~\cite{guo2004}. With 2D Sn$X$ film on different insulating substrates such as CdTe or InAs, the strain from substrate can cause expansion and shrinkage of interlayer spacing, therefore the phonon spectrum can be greatly modulated. The topological property persists in Sn$X$ with lattice constant mismatch from $-7\%$ to $5\%$~\cite{xu2013}. Thus, one could maximize $\lambda_{\text{2d}}$ without changing the topological properties of Sn$X$. In reality, $U_{\text{ph}}$ is usually larger than $V_{\text{ph}}$, and results in a largest energy gain by forming the $s$-wave pairing. However, the effective interaction given by $U$ and $V$ should include effects of the Coulomb interaction and other possible renormalizations. The Coulomb repulsion $U_{\text{coul}}$ and $V_{\text{coul}}$ renormalizes the bare value of $U_{\text{ph}}$ and $V_{\text{ph}}$, respectively. $U_{\text{coul}}\gg V_{\text{coul}}$ due to smaller interorbital wavefunction overlap, and weaker Coulomb screening in 2D makes $U_{\text{coul}}$ larger than that in 3D. The effective interaction parameters are given by $U=U_{\text{ph}}-U_{\text{coul}}$, and $V=V_{\text{ph}}-V_{\text{coul}}$. Therefore the stronger intraorbital repulsion would lead to $U<V$.

\section{Conclusion}

In summary, we have studied topological superconductivity in a doped QSH insulator
and propose 2D doped Sn$X$ as a potential candidate. A wealth of QSH insulating materials could lead to the discovery of the
TRI topological superconductor, which supports the existence of Majorana edge states and Majorana zero energy
modes in vortex cores. We hope the theoretical work here could aid the search for topological superconductor phases in real materials.

\begin{acknowledgments}
This work is supported by the US Department of Energy, Office of Basic Energy Sciences, Division of Materials Sciences and Engineering, under Contract No.~DE-AC02-76SF00515 and the Defense Advanced Research Projects Agency Microsystems
Technology Office, MesoDynamic Architecture Program (MESO) through the Contract No.~N66001-11-1-4105.
\end{acknowledgments}


\begin{thebibliography}{33}
\bibitem [{par()}]{parameters}%
  \BibitemOpen
  \href@noop {} {}\bibinfo {note} {The band structure of SnI is fitted with the
  folloing parameters $A=4.0$~eV$\cdot$\AA, $M_0=0.47$~eV,
  $M_1=-6.0$~eV$\cdot$\AA$^2$, $D_0=-0.4$~eV,
  $D_1=5.2$~eV$\cdot$\AA$^2$.}\BibitemShut {Stop}%
\end{thebibliography}
\end{document}